\documentstyle[12pt]{article}
\newcommand{\leqsim}{\,\raisebox{-0.6ex}{$\buildrel < \over \sim$}\,}
\newcommand{\geqsim}{\,\raisebox{-0.6ex}{$\buildrel > \over \sim$}\,}

\def\etal{{\it et al}\,}
\def\hc{{\rm h.c.}}
\def\ibid{{\it ibid}\,}

\def\P{{\rm Pl}}

\def\gev{\,{\rm GeV}}
\def\mev{\,{\rm MeV}}

\def\tev{\,{\rm TeV}}
\def\kev{\,{\rm KeV}}
\def\ord{{\cal O}}
\def\be{\begin{equation}}
\def\ee{\end{equation}}
\def\ba{\begin{eqnarray}}
\def\ea{\end{eqnarray}}
\begin{document}

\thispagestyle{empty}
\begin{flushright}
{\tt ULB-TH-96/03\\ hep-ph/9603301}
\end{flushright}
\vspace{5mm}
\begin{center}
  {\Large \bf Singlets in Supersymmetry }\\
\vspace{15mm} {\large S.~A.~Abel}\\
\vspace{1cm}
{\small\it Service de Physique Th\'eorique, 
Universit\'e Libre de Bruxelles\\
Boulevard du Triomphe, Bruxelles 1050, Belgium }
\end{center} 
\vspace{2cm}

\begin{abstract}
\noindent It is argued that singlet extensions of the MSSM at the 
weak scale are indicative of either gauged-$R$ symmetry or target
space duality in a string effective action at the Planck scale. The
criteria used are satisfactory primordial nucleosynthesis, absence of
fine-tuning, and absence of cosmological problems such as domain
walls. Models which have a global discrete symmetry such as the NMSSM
may only be accommodated within rather complicated cosmological
scenarios which are also discussed.
\end{abstract}

\pagestyle{plain}
\newpage
\section{Introduction}

There has lately been some interest in the problem of how to
accommodate an extra gauge singlet field into the minimal
supersymmetry standard model (MSSM). This is the simplest
extension which is consistent with a lightest higgs boson whose mass
exceeds the upper bound found in the MSSM~\cite{mssm}. Previously it
was thought that, by acquiring a
vacuum expectation value of $\ord (M_W)$, such a singlet could also
provide a simple solution to a fine-tuning problem in the MSSM,
the so-called `$\mu$--problem'~\cite{muprob,gm}. Because of
difficulties with cosmology (specifically the appearance of domain
walls) this now no longer appears to be the case~\cite{aw,us}. In
fact, it was shown in ref.\cite{us} that models with singlets are likely to
require symmetries in addition to those in the MSSM if they are to
avoid problems with either domain walls or fine-tuning. In this
respect models with gauge singlets are singularly {\em less} efficient
at solving fine-tuning problems. However since they allow for more
complicated higgs phenomenology, it is still worth pursuing them.
This paper concentrates on the task of building an MSSM extended by
a singlet, which avoids reintroducing the hierarchy problem,
fine-tuning, {\em and} domain walls. These three constraints turn out
to place severe restrictions on the type of theory which can be
accommodated at the Planck scale. Conversely the form of the low
energy higgs sector can give important indications of the physics
occuring at the Planck scale.

Let us take as our starting point a low-energy effective theory
which includes all the fields of the MSSM, plus one additional
singlet $N$. The superpotential is assumed to be the standard MSSM
Yukawa couplings plus the higgs interaction
\be
\label{superpot}
W_{\rm higgs}=\mu H_1 H_2 + \mu' N^2 +
\lambda{N}H_{1}H_{2}-\frac{k}{3}N^3,
\ee
and the soft supersymmetry breaking terms are taken to be
of the form  
\begin{eqnarray}
 V_{\rm soft higgs} &=
  &B \mu h_1 h_2 + B' \mu' n^2 +
\lambda A_{\lambda}nh_1h_2- 
\frac{k}{3} A_k n^3 + \hc \nonumber\\
  &&+ m^2_1 |h_1|^2
  + m^2_2 |h_2|^2
  + m^2_N |n|^2,
\end{eqnarray} 
where throughout scalar components will be denoted by lower case
letters.  For the moment let us put aside the question of how the
$\mu$ and $\mu'$ terms get to be so small (i.e. $\ord(M_W)$ instead of
$\ord(M_\P)$), and return to it later. From a low-energy point of view
the only requirement is that the additional singlet should
significantly alter the higgs mass spectrum.  This means that
$\lambda\neq 0$. There are four possibilities which can arise:

If all the other operators are absent, then in the low energy
phenomenology there is an apparent (anomalous) global ${\tilde{U}}(1)$
symmetry (orthogonal to the hypercharge), which leads to a
massless goldstone boson. This situation is briefly reviewed in section 2,
with the conclusion that a significant complication is required in
order that axion bounds are satisfied. In
fact such models can work if the ${\tilde{U}}(1)$ symmetry is only
approximate, being broken to a high order, discrete
symmetry by non-renormalisable terms. In this case an additional 
singlet can provide a solution to the strong CP problem~\cite{graham}.

There are two cases which lead to a discrete symmetry.  These are $\mu
=0$, $k=0$ which leads to a $Z_2$ symmetry, and $\mu =0$, $\mu' =0$
which leads to a $Z_3$ symmetry. The latter is usually 
referred to as the next-to-minimal supersymmetric standard model 
(NMSSM)~\cite{nmssm,ellis}, and will be the main focus of this paper.

Thus the second possibility is that there is an {\em exact} discrete
symmetry, and thus a domain wall problem associated with the existence
of degenerate vacua after the electroweak phase transition. Weak scale
walls cause severe cosmological problems (for example their density
falls as $T^2$ whereas that of radiation falls as $T^4$ so they
eventually dominate and cause power law inflation)~\cite{us}. This is
not true however, if the discrete symmetry is embedded in a broken
gauge symmetry. In this case the degenerate vacua are connected by a
gauge transformation in the full theory~\cite{ls}. After the
electroweak phase transition, one expects a network of domain walls
bounded by cosmic strings to form and then collapse \cite{ls}. (There
are a number of more baroque solutions outlined in ref.\cite{us} which
will not be pursued here.)  In addition there is no gauge singlet to
destabilise the gauge hierarchy~\cite{destab}. This
situation is examined in section 3, where it is found that bounds 
from primordial
nucleosynthesis (essentially on the reheat temperature after
inflation) require that the potential be very flat.  Such potentials
occur naturally as the low energy approximation to string theory and
two examples will be examined. The first, involving an extra $U(1)_X$ gauge
symmetry has difficulty satisfying anomaly cancellation
conditions. The second example, in which the $Z_3$ is embedded in
$SU(3)^3$ looks more promising, although the presence of the $N^3$
term necessitates large multiplets which do not exist in the simpler
string models. In addition this mechanism depends rather strongly on
the cosmology which is not yet fully known for string effective
actions, and so models with discrete symmetry (such as
the NMSSM) remain questionable.

The third possibility is that the discrete symmetry is
broken~\cite{zko} by gravitationally suppressed
interactions~\cite{ellis,rai}.  This was the case considered and
rejected in ref.\cite{us}.  Here the very slight non-degeneracy in the
vacua, causes the true vacuum to dominate once the typical curvature
scale of the domain wall structure becomes large enough. However one
must ensure that the domain walls disappear before the onset of
nucleosynthesis and this means that the gravitationally suppressed
terms must be of order five. It was shown in ref.\cite{us} that, no
matter how complicated the full theory (i.e. including gravity), there
is {\em no} symmetry which can allow one of these terms, whilst
forbidding the operator $\nu N$, where $\nu$ is an effective
coupling. Furthermore, any such operator large enough to make the
domain walls disappear before nucleosynthesis generates these terms at
one loop anyway (with magnitude $\sim M_W^2 M_\P N$), even if they are
set to zero initially. This constitutes a reintroduction of the
hierarchy problem as emphasised in ref.\cite{destab}.

The final case which is discussed in section 4, is when there is no
discrete symmetry at the weak scale (exact or apparent). This is true
when either $\mu\neq 0$ or both $\mu'\neq 0$ and $k\neq 0$. Here the
arguments of ref.\cite{us} will again imply fine-tuning, {\em except}
in the case that there is either a target space duality symmetry,
again coming from the low energy approximation to string theory or an
$R$-symmetry at the Planck scale. For the reasons discussed in
ref.\cite{herbi}, gauged $R$-symmetry~\cite{herbi,gaugedr} might be
favoured over global, although the arguments presented will apply to
either case.  In these cases, fine tuning can be
avoided since the transformation of the superpotential is different
from that of the K\"ahler potential. Thanks to the experimental
signatures discussed in ref.\cite{herbi}, the case of gauged
$R$-symmetry should be easily distinguishable. It is also established
that the higher loop contributions to the effective potential (in the
framework of $N=1$ supergravity), do not destabilise the gauge 
hierarchy either.

Thus it can be argued quite generally that, provided one considers lack
of fine-tuning to be a legitimate constraint, gauge singlets at the weak
scale are likely to be embedded in either a string effective action, or
a gauged $R$-symmetry. It should be stressed that this is not 
simply a restatement
of the conclusion drawn in ref.\cite{destab}. There the absence of
destabilising divergences implied that there are no singlets in the
full theory including gravity. In fact {\em any} symmetry (such as the
discrete $Z_3$ symmetry in the NMSSM) is sufficient to ensure this, as
was demonstrated earlier.

\section{The Global $\tilde{U}(1)$ Case}

Models in which $\lambda{N}H_{1}H_{2}$ is the only operator appearing
in the higgs superpotential have a continuous $\tilde{U}(1)$ symmetry
which is both global and anomalous. This section merely reviews
this situation, which is unworkable without the addition of extra
suppressed terms, since the model fails to satisfy axion bounds. One
may rectify this by adding gravitationally suppressed terms breaking
the $\tilde{U}(1)$ symmetry to a high-order discrete symmetry. In this
case, for the (pseudo) goldstone boson to provide a solution to the strong
CP problem, an additional singlet is required.

The quantum numbers of the higgs fields may be taken as follows;
\vspace{0.5cm}
\begin{center}
\begin{tabular}{||l|r|r||}   \hline
           & $U(1)_Y$ & $\tilde{U}(1)$ \\ \hline\hline 
$N$     & $0 $     & $ -2$ \\ \hline 
$H_1$  & $1 $     & $ 1$ \\ \hline 
$H_2$  & $-1$     & $ 1$ \\ \hline \hline
\end{tabular}
\end{center}
\vspace{0.5cm}
with the right and left-handed quarks transforming appropriately to
keep the Yukawa terms invariant. The anomaly may be expressed as the
non-conservation of the current generated under such a rotation:
\begin{equation} 
\label{anom}
\partial_\mu \tilde{J}^\mu = 3 \theta_N
\frac{F_{\mu\nu}\tilde{F}^{\mu\nu}}{32 \pi^2}
\end{equation}
where $N\rightarrow \exp (i \theta_N) N$.  There are a number of well
rehearsed problems which now arise, and it is worth examining them in
some more detail for this case.

The first problem is due to the appearance of domain
walls. Eq.(\ref{anom}) tells us that the partition function is
invariant when $\theta_N=0,\pm 2\pi /3$, and that there are
therefore three degenerate vacua after the $\tilde{U}(1)$ symmetry
has been broken by QCD instantons. Domain walls appear across which
$\theta_N$ changes by $\pm 2\pi /3$. In fact the residual $Z_3$
symmetry is precisely the anomaly free discrete symmetry of the
NMSSM. This difficulty could potentially be resolved by the
Lazarides-Shafi mechanism~\cite{ls} (i.e. embedding the discrete
symmetry in a gauged continuous symmetry) which will be discussed 
later in the context of the NMSSM.

However a second and more serious problem is the existence of an axion
associated with the $\tilde{U}(1)$ symmetry.  In fact this model is
just a particular case of the $DFSZ$ axion model~\cite{DFSZ} except
here one is hampered considerably by the fact that there is only one
parameter, namely $\lambda$, which can be adjusted. 
Anticipating that the anomalous $\tilde{U}(1)$ can provide a
realisation of the Peccei-Quinn mechanism, in $DFSZ$ models the
mass of the axion is given by
\begin{equation}
m_a=\frac{\sqrt{m_u m_d}}{m_u+m_d} \frac{m_\pi f_\pi}{f_a} ,
\end{equation}
where $f_a$ is the axion decay constant~\cite{CPrefs}.  
Astrophysics and cosmological considerations place lower and upper bounds
respectively on the value of $f_a$~\cite{CPrefs};
\begin{equation}
10^9 \leqsim f_a \leqsim 3\times 10^{12} \gev.
\end{equation}
Since $\langle h^0_1 \rangle $ and $\langle h^0_2 \rangle $ are
constrained to be of order $M_W$, this means that $f_a \approx
\langle n \rangle$ (where lower case letters denote
scalar components). Thus one requires the parameter $\lambda
\leqsim 10^{-6}$ for suitable electroweak symmetry breaking to
occur (taking $\lambda |n|<1\tev$). On the other hand the scalar
potential is of the form
\begin{eqnarray}
 V_{\rm neutral-scalars} &=
  &B \lambda^2 \left( |n h^0_1|^2+|n h_2^0|^2+|h_1^0 h_2^0|^2\right) +
\lambda A_{\lambda}nh^0_1h^0_2+ \hc \nonumber\\
  &&+ m^2_1 |h_1^0|^2
  + m^2_2 |h_{2}^0|^2
  + m^2_N |n|^2.
\end{eqnarray} 
Because the soft supersymmetry breaking scalar masses and trilinear
terms are expected to be of order $M_W$, one cannot expect $\langle n
\rangle$ to be much more than $\lambda M_W$. Thus $f_a\geqsim
10^{9}\gev$ is incompatible with electroweak symmetry breaking.

Plainly some extension to the superpotential is required, and, at the
risk of complicating things considerably, one might consider adding
undetectable, non-renormalisable terms such as $N^m /M_\P^{m-3}$ to
the superpotential.  Now a minimum {\em can} appear for $\langle n
\rangle \sim M_\P (M_W/ M_\P)^{1/(m-2)}$, lying roughly in the desired
range if $m=4$ or 5 (where $M_\P$ implies the reduced Planck mass). Also
there is now no domain wall problem, since the remaining discrete
symmetry (i.e. a $Z_m$ symmetry) is anomalous. This model with one
singlet does not resolve the strong CP problem however.  This is
because the effective potential in $\overline{\theta}=\theta_{QCD} +3
\theta_N + \mbox{arg det}(\lambda_u \lambda_d)$ which is of the form
(see Cheng in ref.\cite{CPrefs} and references therein),
\begin{equation}
V_{\rm PQ}=m_a^2 f_a^2 (1-\cos (\overline{\theta})),
\end{equation}
is outweighed by the dependence of the potential on $\theta_N$, 
\begin{equation}
V_{\rm PQ-breaking}=-\frac{M_W}{M_\P^{m-3}} |n|^m \cos (m \theta_N)
\end{equation}
which drives $\theta_N$ to $0$, $2 \pi /m$, $4 \pi/m \ldots $ or $2
\pi $, rather than the value which gives $\overline{\theta}=0$. There
is then no hope of using the Peccei-Quinn mechanism to solve the
strong CP problem.

This situation is similar to one which has been studied in some detail
by Casas and Ross~\cite{graham}.  With the aid of discrete $Z_2\times
Z_3$ symmetry and an extra singlet ($M$), they constructed models in
which the terms responsible for breaking the global $\tilde{U}(1)$
symmetry were of dimension-12. This gives a successful implementation
of the (pseudo) Peccei-Quinn mechanism with a suitably low value for 
$\lambda$. 

So the case of continuous, global $\tilde{U}(1)$ seems potentially
promising, if the symmetry is only the approximate, low energy
manifestation of a high order, anomalous, discrete symmetry. However,
the question of successful (dynamical) realisation of electroweak
symmetry breaking has not yet been addressed for these models.   

\section{The Discrete Symmetry Case}

This section demonstrates how models with discrete symmetry may be
embedded in a gauge symmetry which is spontaneously broken at some
high scale. In this case the apparently distinct degenerate vacua of
the low energy theory are connected by a gauge transformation.  At all
times the main aim is to keep the model as `minimal' as possible,
however it will become apparent that constraints coming from
primordial nucleosynthesis require that the potential to be very flat,
leading us in the following section to consider string effective 
supergravity models.

This scenario was suggested by Lazarides and Shafi~\cite{ls}, and for
definiteness (and simplicity) first consider adding an extra $U(1)_X$
gauge symmetry to the the NMSSM, (so that $\mu=0$ and $\mu'=0$), which
is spontaneously broken at some high scale. The discrete symmetry is
an apparent $Z_3$ symmetry when every chiral superfield is rotated by
$e^{2\pi i/3}$ (i.e. only trilinear terms appear in the
superpotential). This rotation may be embedded in $U(1)_X$ by
introducing an additional singlet field $\Phi$ which generates the
superpotential as effective interactions via non-renormalisable terms,
\be 
\label{nmssmpot}
W_{\rm higgs}=\lambda \frac{\Phi}{M_\P}N H_{1}H_{2}-
\frac{k}{3}\frac{\Phi}{M_\P}N^3.
\ee
The field $\Phi $ will acquire a VEV of order $M_\P$, forming
strings as it does so.  The latter involve physics at distances 
much smaller than the weak scale. Because of this it is useful to 
think in terms of the
previous $\tilde{U}(1)$ rotation as an additional {\em approximate}
global symmetry operating orthogonally to the two gauge symmetries
since it does not involve $\Phi $ (becoming exact when $k=0$).
The quantum numbers of the higgs fields are as follows;
\vspace{0.5cm}
\begin{center}
\begin{tabular}{||l|r|r|r||}   \hline
       & $ U(1)_X$     & $U(1)_Y$ & $\tilde{U}(1)$ \\ \hline\hline 
$N$    & $1      $ & $0 $     & $-2          $ \\ \hline 
$H_1$  & $1      $ & $1 $     & $ 1          $ \\ \hline 
$H_2$  & $1      $ & $-1$     & $ 1          $ \\ \hline 
$\Phi$ & $-3     $ & $0 $     & $ 0          $ \\ \hline\hline
\end{tabular}
\end{center}
\vspace{0.5cm}
The global $\tilde{U}(1)$ symmetry and local $U(1)_X$ symmetry have a
non-trivial intersection which is precisely the $Z_3$ operation
defined above. One is forced to involve the two higgs fields as well
as $N$, which complicates the job of anomaly cancellation. For the
moment however this question can be neglected, since the physics we are
presently concerned with is purely classical. Eventually anomalies
will be cancelled by assigning non-zero $U(1)_X$ numbers
to the remaining fields in the visible sector.

Now consider what happens when these symmetries are spontaneously
broken down to $SU(3)_c\times U(1)_{em}$.  Obviously for the effective
low-energy theory to resemble the NMSSM, this must happen with
$\langle |\phi |\rangle=\rho_\phi=\ord (M_\P)$. For the moment
therefore, let us simply assume that at some point a Lagrangian
develops for the scalar component ($\phi$) of the $\Phi $ superfield
of the form,
\be
\label{philag}
{\cal{L}}_\phi = ({\cal{D}} \phi)^{\dagger} {\cal{D}} \phi + M^2
|\phi |^2 - L |\phi |^4
\ee
where one would {\em a priori} expect $| M |=\ord (M_\P) $. When $M^2
> 0$, the potential turns over at the origin, developing a minimum at
$\langle |\phi| \rangle=\rho_\phi=\sqrt{M^2 /2 L }$.  The goldstone 
mode can be identified as the field coupling linearly to the $X^\mu$
current;
\be
G_\mu = (3 \sqrt{2}g_X /M_X)\left( \rho_\phi^2\partial_\mu\theta_\phi- 
\frac{1}{3}\left(  
\rho^2_n\partial_\mu\theta_n+\rho_1^2\partial_\mu\theta_1
+\rho_2^2\partial_\mu\theta_2\right)\right)
\ee
where, 
\be
M_X^2 = 18 g_X^2 \left( \rho_\phi^2 + \frac{1}{9} \left(
    \rho_n^2+\rho_1^2 +\rho_2^2\right)\right),
\ee
and where the scalar components are defined such that, $\phi = |\phi|
e^{i \theta_\phi}$, $n=|n| e^{i\theta_n}$, $h_1=|h_1| e^{i \theta_1}$,
$h_2=|h_2| e^{i \theta_2}$, and $\langle |n|\rangle = \rho_n $,
$\langle |h_1|\rangle = \rho_1 $, $\langle |h_2|\rangle = \rho_2 $.
Once the weak scale higgs fields have acquired a vacuum expectation
value, there is therefore a tiny amount of mixing between the heavy
and light bosons. In fact defining the usual mass of the $Z$ by $M_Z^2
= (g_y^2+g_2^2)(\rho_1^2+\rho_2^2)/2 $, and also $m^2= g_X
\sqrt{g_y^2+g_2^2}(\rho_1^2-\rho_2^2)$, one finds that the mass 
eigenstates are given by
\ba
Z^{'\mu}&=&\cos\theta_X Z^\mu-\sin\theta_X X^\mu
\approx Z^\mu-\frac{m^2}{2 M_X^2}X^\mu\nonumber\\
X^{'\mu}&=&\sin\theta_X Z^\mu+\cos\theta_X X^\mu
\approx X^\mu+\frac{m^2}{2 M_X^2}Z^\mu,
\ea
where 
\be
\theta_X=\frac{1}{2}\tan^{-1}\left(\frac{m^2}{M_X^2-M_Z^2}\right).
\ee
The mass of the $Z$ is virtually unchanged;
\be
M^{2}_{Z'}=M^2_Z-3 \frac{m^2}{M_X^2} m^2.
\ee

To describe the topological defects that result from this pattern of
breaking, consider first what happens after the first stage, 
once the $U(1)_X$ symmetry is spontaneously broken. Vortex 
solutions appear of the form  
\ba
\mbox{Lim}_{r\gg M_X^{-1}}\langle \phi \rangle 
&=& \rho_\phi e^{-i \alpha (\theta)} \nonumber\\
\mbox{Lim}_{r\gg M_X^{-1}}\langle X_\mu \rangle 
&=& \frac{1}{3 g_X}\partial_\mu \alpha (\theta)    
\ea
with winding number $s=3 t + u$ where $u=\{0,1,2\}$, $t$ is an integer, 
and $\alpha (\theta)$ is a continuous function of the azimuthal angle
$\theta$ such that 
\be
\alpha(\theta+2\pi)=\alpha (\theta) + 2 s \pi 
\ee
(assuming translational symmetry in the $z$ direction). 
This solution corresponds to a gauged $U(1)_X$ rotation through 
$2 s \pi /3$ as one goes around the
string and therefore has zero energy density as $r\rightarrow \infty$.
The integer $u$, labels the elements of $\pi_1(G/H)=Z_3 $. As one
parallel transports a test particle, $i$, around a string it picks 
up a phase of $\exp (2\pi i u X_i /3)$. 

Now consider what happens around a string at the second stage of
symmetry breaking in which the $n$, $h_1$ and $h_2$ fields acquire a
VEV. The VEVs are described by functions which are single valued, 
such that the $\theta $ dependence of the
condensates obeys $\langle n (\theta+2\pi)\rangle$, 
$\langle h_1 (\theta+2\pi)\rangle $, 
$\langle h_2 (\theta+2\pi)\rangle  = 
\langle n (\theta)\rangle $, $\langle h_1 (\theta)\rangle$, 
$\langle h_2 (\theta)\rangle $. 
They may be written as the sum of the above $U(1)_X$ rotation 
and an additional global ${\tilde U}(1)$ rotation through 
$-2 \tilde{s} \pi /3$, so that for the remaining higgs fields,
\ba
\mbox{Lim}_{r\rightarrow\infty}\langle n \rangle 
&=& \rho_n (r) e^{i (\alpha(\theta)+2\tilde{\alpha}(\theta)) /3} \nonumber\\
\mbox{Lim}_{r\rightarrow\infty}\langle h_1 \rangle 
&=& \rho_1 (r) e^{i (\alpha(\theta)-\tilde{\alpha}(\theta)) /3} \nonumber\\
\mbox{Lim}_{r\rightarrow\infty}\langle h_2 \rangle 
&=& \rho_2 (r) e^{i (\alpha(\theta)-\tilde{\alpha}(\theta)) /3},
\ea
where 
\be
\label{cond}
\tilde{\alpha}(\theta+2\pi)=\tilde{\alpha}(\theta) + 2 \tilde{s} \pi ,
\ee
and $\tilde{s}=3 \tilde{t} +u$ (same $u$). This combination of gauge
and global rotations is similar to the {\em hybrid} string
configuration described in ref.~\cite{ls}
(except here of course the global symmetry is already explicitly
broken at tree level). By simple scaling arguments, one expects the string to
have a width of $\ord (L^{-1/2} \rho_\phi^{-1} )$ and a mass per
unit length of $\ord (\rho_\phi^2)$. 

This description is useful, since (as $r\rightarrow
\infty $), the $\alpha(\theta)$ dependence may be taken out of
the low energy (that is near $M_W$) effective Lagrangian, 
by expressing the theory around the string in 
terms of the constants $\hat{\lambda}=\lambda \rho_\phi/M_\P$ and 
$\hat{k}=k \rho_\phi/M_\P$, and the superfields 
\ba
\hat{N}
&=& N e^{-i \alpha (\theta) /3} \nonumber\\
\hat{H}_1  
&=& H_1 e^{-i \alpha (\theta) /3} \nonumber\\
\hat{H}_2
&=& H_2 e^{-i \alpha (\theta) /3}.
\ea
The effective Lagrangian in terms of the roofed parameters is simply
the NMSSM. Close to the string for $M_W^{-1}\gg r \gg L^{-1/2}
\rho_\phi^{-1}$, the gradient energy density ($\sim \rho^2/r^2$)
dominates over the potential energy ($\sim \rho^4$).  Here the scalar
fields approximately obey the Laplace equation, with solutions,
\ba
\langle \hat{n} \rangle 
&=& \rho_n \sum_{l} \left( A_n^l z^{l+2u/3}+B_n^l {\overline{z}}^{l-2u/3} 
\right) \nonumber\\ \langle \hat{h}_1 \rangle 
&=& \rho_1 \sum_{l} \left( A_1^l z^{l-u/3}+B_1^l {\overline{z}}^{l+u/3}   
\right) \nonumber\\ \langle \hat{h}_1 \rangle 
&=& \rho_2 \sum_{l} \left( A_2^l z^{l-u/3}+B_2^l
  {\overline{z}}^{l+u/3}
\right) ,
\ea
where $l$ is an integer and $z=(r/r_0) e^{i\theta}$. 

On larger scales, $r\geqsim\ord (M_W^{-1})$, the potential energy
density becomes important.  If $u=0$ (i.e. when the winding number is
divisible by three) the minimum energy solution for $r\gg M_W^{-1}$ is
simply constant $\langle \hat{n} \rangle$, $\langle \hat{h}_1 \rangle$ and
$\langle \hat{h}_1 \rangle$, which can be trivially matched onto the $l=0$
solution above. This case is just the usual cosmic string of
mass/unit length $\eta = \ord (\rho_\phi^{2})$. But if $u=\pm 1$ the minimum
energy solution is a single wall of exactly the same form as in ref.\cite{us}, 
which has thickness $\sim M_W^{-1}$, and across which the fields
acquire a phase
\be
\langle \hat{n} \rangle \rightarrow e^{4 u i\pi /3}
\langle\hat{n}\rangle \mbox{ ; } 
\langle \hat{h}_1\rangle\rightarrow e^{-2 u i\pi /3}
\langle\hat{h}_1\rangle \mbox{ ; } 
\langle \hat{h}_2\rangle\rightarrow e^{-2 u i\pi /3}
\langle\hat{h}_2\rangle .
\ee 
The configuration in this case is a domain wall of mass/unit area
$\sigma = \ord (M_W^3)$, bounded by a cosmic string of mass/unit
length $\eta = \ord (\rho_\phi^{2})$. 

The cosmology of this type of string/wall system was discussed in 
refs.\cite{ls,wallevol}. At temperatures high above the electroweak 
phase transition but below $M_X$, the 
strings evolve freely, straightening under their own tension until
there is roughly one string per horizon. Once the temperature drops
below the electroweak phase transition some of them become connected 
by domain walls. These also evolve until there is roughly one domain 
wall bounded by a string per horizon, at which point the system 
collapses under its own tension~\cite{ls,wallevol}. 

Nucleosynthesis imposes some simple bounds on the parameters
$M$ and $L$ appearing in eq.(\ref{philag}) as follows. Firstly it
should be stressed that since $M_X$ is much larger than $M_W$, the 
domain walls are stable within the lifetime of the universe. This is 
because the probability for quantum tunnelling a hole bounded by a
string in any wall is suppressed by the Boltzmann factor $\exp
(-M^4_X/M_W^3 T)$\cite{ls}. (Interestingly, in addition to the usual 
configuration where two strings
appear with opposite winding number giving a hole, there is also the 
possibility for a three wall vertex to collapse by inserting into the 
vertex two strings of winding number $+1$ and one of winding number
$-2$.) This means that any period of inflation cannot have inflated 
away the strings. The creation and subsequent decay of 
gravitinos places an upper limit on the reheat temperature after
inflation $T_R \leqsim 10^{9}\gev$~\cite{cosmobounds}. 
Thus the $U(1)_X$ symmetry must be restored at temperatures 
below $T_R$, giving 
\be
\label{requirement}
T_c=\sqrt{48 M^2/(16 L + 27 g_X^2)}\leqsim 10^{9}\gev,
\ee
in a high-temperature mean-field approximation. (Note that the 
temperature in the $\Phi$ sector need not be the same as in the 
visible sector since they may not be in equilibrium, however the 
bound on $T_R$ applies to all sectors separately.) This bound is avoided
if the gravitino is very heavy (heavier than $50\tev $) or extremely 
light (lighter than $1\kev $), but neither of these possibilities 
may be easily realised (see Sarkar in ref.\cite{cosmobounds} and
references therein). The condition in eq.(\ref{requirement}) 
is in conflict with the phenomenological requirement that 
$\rho_\phi \sim M_\P$, since together they imply that
\ba
L &\approx & \frac{9 g^2_X T^2_c}{32 M^2_\P} \leqsim
10^{-20}\nonumber\\
M &\approx & \frac{3}{4} g_x T_c \leqsim 10^{-10} M_\P,
\ea
assuming that $g_X = \ord (1)$. This flatness in the potential
looks very unnatural unless it is enforced by some symmetry, at least in
the context of field theory. However string effective supergravity
models naturally have flat directions, and so the next section
considers how the mechanism might, in principle, be made to work in these.

\section{The Lazarides-Shafi Mechanism in String Theory}

Generally, string effective supergravity has many flat directions, 
some of which correspond to moduli determining the size and shape of the
compactified space. Furthermore these moduli have discrete 
duality symmetries, which at certain points of enhanced symmetry become
continuous gauge symmetries~\cite{duality}. 
Thus one would expect the Lazarides-Shafi
mechanism to be directly applicable here. (This possibility has
also been alluded to in refs.\cite{david1,david2}.) In this section 
is shown
that this is indeed the case, although anomalies and again
nucleosynthesis bounds, force the resulting models to be rather
clumsy. In fact anomaly cancellation virtually eliminates $U(1)_X$ 
as a reasonable symmetry in which to embed $Z_3$. More generally one
expects the cosmology to be rather troublesome. 

In Calabi-Yau models, abelian orbifolds and fermionic strings the
moduli include three K\"ahler class moduli ($T$-type) which are always
present, plus the possible deformations of the complex structure
($U$-type), all of which are gauge singlets. Additionally there
will generally be complex Wilson line fields~\cite{moduli,moduli2}.
When the latter acquire a vacuum expectation value they result in
the breaking of gauge symmetries. There has been continued interest
in string effective actions since they may induce the higgs
$\mu$-term~\cite{gm,moduli2,ant1,brignole}, be able to explain the
Yukawa structure~\cite{kpz,binetruy}, and be able to explain the
smallness of the cosmological constant in a {\em no-scale}
fashion~\cite{kpz,noscale}. Since the main objective here is 
simply to find a route to a low energy model with visible 
higgs singlets and apparent discrete symmetry, these questions
will only be partially addressed.

Typically the moduli and matter fields describe a space whose local 
structure is given by a 
direct product of $SU(n,m)/SU(n)\times SU(m)$ and $SO(n,m)/SO(n)\times
SO(m)$ factors~\cite{moduli,moduli2}. As an example consider
the K\"ahler potential derived in refs.\cite{moduli2}, which 
at the tree level is of the form 
\be
\label{stringkahler}
K=-\log (S+\overline{S}) -\log
[(T+\overline{T})(U+\overline{U})-\frac{1}{2}
(\Phi_1+\overline{\Phi}_2)(\Phi_2+\overline{\Phi}_1)]+\ldots
\ee
The $S$ superfield is the dilaton/axion chiral multiplet, and the
ellipsis stands for terms involving the matter fields. 
The fields $\Phi_1$ and $\Phi_2$ are two Wilson line moduli. In
ref.\cite{brignole}, these fields were identified with the neutral
components of the higgs doublets in order to break electroweak
symmetry, and also to provide a $\mu$-term. Here however, their role 
is to break $U(1)_X$, and so they are instead chosen to have 
X-charges of $-1$ and $+1$
respectively. Problems such as how the dilaton acquires a VEV, 
or the eventual mechanism which seeds supersymmetry breaking will not
be addressed here. 

The moduli space is given locally by 
\be
{\cal{K}}_0 = \frac{SU(1,1)}{U(1)}\times \frac{SO(2,4)}{SO(2)\times
  SO(4)},
\ee
which ensures the vanishing of the scalar potential at least at the 
tree level, provided that the $S$, $T$ and $U$ fields all participate
in supersymmetry breaking (i.e. $G_S$, $G_T$, $G_U\neq 0$). 
In fact writing the K\"ahler function as
\be
G=K(z,\overline{z})+\ln \left| W(z) \right|^2 ,
\ee
where $z_i$ are generic chiral superfields, the scalar potential becomes 
\be
V_s = - e^{G} \left(3- G_i G^{i\overline{j}} G_{\overline{j}} \right)
+ \frac{g^2}{2} {\rm Re} (G^i T^{Aj}_iz_j)(G^k T^{Al}_kz_l),
\ee
where $G_i=\partial G/\partial z_i$, and $G^{i{\overline{j}}}=
(G_{{\overline{j}}i})^{-1}$. The dilaton contribution separates, and
gives $G_S G^{S\overline{S}} G_{\overline{S}}=1$. To show that the
remaining contribution is $2$, it is simplest to define the vector
\be
A^\alpha= a (t,u,h,\overline{h})
\ee
where the components are defined as $\alpha=(1\ldots 4) \equiv
(T,U,\Phi_1,\Phi_2)$, and $u=U+\overline{U}$, $t=T+\overline{T}$,
$h=\Phi_1+\overline{\Phi}_2$. It is easy to show that
\be
G_\alpha A^\alpha =-2 a.
\ee
The vector $A^\alpha$ is designed so that $G_{\overline{\beta}\alpha} 
A^\alpha $ is proportional to $G_{\overline{\beta}}$; viz,
\be
G_{\overline{\beta}\alpha}A^\alpha = - a G_{\overline{\beta}}.
\ee 
Multiplying both sides by $G_\alpha G^{\alpha\overline{\beta}}$ gives
the desired result, i.e. that $G_\alpha G^{\alpha\overline{\beta}}
G_{\overline{\beta}}=2$. Thus, if the VEVs of the matter fields are
zero, the potential vanishes and is flat for all values of the moduli
$T$ and $U$, along the direction $\langle |\Phi_1|\rangle = \langle
|\Phi_2|\rangle=\rho_{\phi}$ (since this is the direction in which the
$D$-terms vanish).  The gravitino mass is therefore undetermined 
at tree level, being given by
\be
m_{3/2}^2 = \langle e^G \rangle = \frac{|W|^2}{s(ut-2 \rho^2_{\phi})}.
\ee
In addition to the properties described above, there is an 
$O(2,4,Z)$ duality corresponding to automorphisms of the compactification 
lattice~\cite{duality,moduli2}. This constrains the possible form of the
superpotential. The $PSL(2,Z)_T$ subgroup implies 
invariance under the transformations~\cite{duality,moduli2},  
\ba
\label{dualtrans}
T &\rightarrow& \frac{aT-ib}{icT+d} \nonumber\\
U &\rightarrow& U-\frac{ic}{2}\frac{\Phi_1\Phi_2}{icT+d} \nonumber\\
\Phi_i &\rightarrow& \Phi_i (icT +d)^{n_i},
\ea
where $a,b,c,d~\epsilon~Z$, $ad-bc=1$, and 
where $\Phi_i$ stands for general matter superfields with
weight $n_i$ under the modular transformation above. 
The $\Phi_1$ and $\Phi_2$ fields have modular weight $-1$. 
It is easy to verify the invariance of the K\"ahler function under
this transformation provided that 
\be
\label{wdual}
W\rightarrow (ic T + d)^{-1} W.
\ee
The superpotential should be defined to be consistent with this
requirement in addition to X-charge invariance, and this leads to a
constraint on the modular weights of the Yukawa couplings and matter
fields. Anomalies occur here also, and must be cancelled in addition
to the gauge anomalies.

Now let us try to include the Lazarides-Shafi mechanism by assuming 
that the superpotential contains a higgs portion,
\be 
W_{\rm higgs}=\lambda
\frac{\Phi_1^{u_1}\Phi_2^{u_2}}{M^{u_1+u_2}_\P}
N H_{1}H_{2}-
\frac{k}{3}
\frac{\Phi_1^{w_1}\Phi_2^{w_2}}{M^{u_1+u_2}_\P}
N^3,
\ee
where, for the purposes of anomaly cancellation, the powers are as
general as possible. In addition let the other Yukawa 
couplings have factors of powers of $\Phi_1$ and $\Phi_2$,
\be
W_{\rm yukawa}=\lambda_{ijk}
\left(\frac{\Phi_1^{l_{ijk}}}{M^{l_{ijk}}_\P}\right)^{m_{ijk}}
\left(\frac{\Phi_2^{k_{ijk}}}{M^{k_{ijk}}_\P}\right)^{l_{ijk}}
z^iz^jz^k = \hat{\lambda}_{ijk}z^iz^jz^k,
\ee
where from now on $z^i$ should be understood as a generic {\em
visible} sector chiral superfield.  The Yukawa couplings will
eventually absorb factors of $K^{-1/2}$ when the physical scalars are
correctly normalised.  The low energy Yukawa couplings are then
functions of the VEVs of $S$, $T$, $U$, $\Phi_1$ and $\Phi_2$ fields
(and possibly other fields which are uncharged under $U(1)_X$), but
they are effectively constants in the MSSM, once the
hidden sector has been decoupled.

The cosmology of string actions is still being developed, but let
us adhere closely to the current thinking (which has been
summarized by Lyth and Stewart in ref.\cite{david2}).  In order
for the strings to be formed, the $U(1)_X$ symmetry must at some
point be restored after inflation with $|\Phi | \ll M_\P $. This
is possible if the $X$ gauge boson is in thermal equilibrium
after inflation, the flat direction being lifted away from the
origin by thermal mass terms $\sim |\Phi|^2 T^2$ ($T$ being the
temperature here). This term holds the $\Phi_i$ fields (here
playing the role of `flatons') at the origin, until the
temperature drops below $m_{3/2}$. Thus, during the period where
$\sqrt{m_{3/2} \rho_\phi} \geqsim T \geqsim m_{3/2}$, there is
`thermal' inflation of only a few $e$-folds (10 or so) due to the
non-zero vacuum energy density.  It is assumed of course that the
vacuum energy density is zero where $|\Phi |$ eventually
obtains its VEV (with $\langle |\Phi | \rangle = \rho_\phi \sim M_\P$). 
After inflation the 
energy is converted into oscillations of the $\Phi_i$ fields, which 
eventually decay with a reheat temperature
\be
T_D \sim \left(\frac{10^{11} \gev}{\rho_\phi } 
\right)\gev .   
\ee
This gives rise to the cosmological `moduli'
problem~\cite{subir,david1,david2}.  Successful nucleosynthesis requires
that $T_D\geqsim 10\mev $ implying that $ \rho_\phi \leqsim 10^{14}
\gev$. (Note that the possibility that the minimum remain fairly
constant is excluded here by the requirement that $|\Phi |\ll
M_\P$ initially and by the form of the Yukawa couplings which
require $|\Phi | \sim M_\P$ eventually; there is bound to be a
large amount of entropy stored up in moduli oscillations.) The only
solution appears to be for a second period of thermal inflation to
occur which dilutes the abundance of the $\Phi_i$ particles after
they have thermalized, involving a second flaton whose VEV should
be less than $10^{12} \gev $~\cite{david2}.

A second difficulty, effectively rules out the $U(1)_X$ case entirely,
when one considers anomaly cancellation and duality invariance in the
effective action.  The anomaly cancellation conditions (which may be
found in ref.\cite{dps} for example) must be taken to be
$\frac{3}{5}A_1=A_2=A_3=A'_1=0$.  Since the $\Phi_i$
particles are required to be in equilibrium, one should avoid using 
the Green-Schwarz
mechanism to help cancel these~\cite{greenschwarz}. This is because
Green-Schwarz anomaly cancellation involves a Fayet-Illiopoulos term
in the potential at the Planck scale, which gives the $X$ gauge-boson
a mass of ${\cal O}( M_\P )$~\cite{dsw}.  (For the same reason it is
not possible to choose the $U(1)_X$ symmetry to be a gauged
$R$-symmetry.) This provides severe restrictions on the allowed Yukawa
couplings. Since there is a $U(1)_X$ symmetry, using $\rho_\phi$ to
determine the fermion masses and mixings as in ref.\cite{dps}, would
seem a possibility. However one must be careful not to induce
flavour changing neutral currents in the low-energy theory. This can
be ensured by assuming that both the modular weights (denoted by $n$)
and $X$-charges (denoted by $x$) are generation independent. In
addition one can allow (and in fact this will be necessary) a
generation-degenerate weight for each of the Yukawa couplings. Thus
the best one could hope for here is to explain the hierachy between
$\hat{\lambda}_t$ and $\hat{\lambda}_b$, $\hat{\lambda}_\tau$. The
embedding of the $Z_3$ symmetry in $U(1)_X$ requires that
$x_{h_1}=x_{h_2}=x_n$.  First one may solve for the gauge anomalies
eliminating $x_q$, $x_u$ and $x_l$. The quadratic $A'_1$ anomaly is then
linear in $x_e$ which is also eliminated. Since the cubic
and gravitational anomalies may be cancelled with some extra
hidden sector fields, they will not be considered.  
Next the condition 
eq.(\ref{wdual}) may be solved, together with the corresponding 
anomaly cancellation;
$B_1=B_2=B_3$ using the notation of ref.\cite{binetruy}. Assuming that
the top quark mass is not suppressed by very many powers of $\rho_\phi$, one
then finds that the Yukawa couplings must themselves have a modular
weight. Finally one can solve for X-charge invariance of the Yukawa
interactions and
eliminate $x_{h_1}$, $x_d$ and the $X$-charge of $\hat{\lambda}_\tau $. 
This leaves the freedom
to choose the weights of $\lambda_t $, $\lambda_b$, $\lambda_\tau $ 
and $k$ (which for example may be done so that $B_1=B_2=B_3=0$), and 
the X-charges of $\hat{\lambda}_t$ and $\hat{\lambda}_b$, together with
$x_{\Phi_1}=-x_{\Phi_2}$, $u_2$, $w_2$, $n_{h_1}$ and $n_q$. The 
charges and weights may for instance be chosen as follows,
\vspace{0.5cm}
\begin{center}
\begin{tabular}{||l|r|r||}   \hline
               & $x_i$     & $n_i$    \\ \hline\hline 
$N$            & 3        & 3      \\ \hline 
$H_1$          & 3        & 1        \\ \hline 
$H_2$          & 3        & 5     \\ \hline 
$\Phi_1$       & -1        & -1       \\ \hline 
$\Phi_2$       &  1        & -1       \\ \hline 
$Q_L$          & -1/3     & -1       \\ \hline 
$U^c_R$        & -5/3     & -7/3      \\ \hline 
$D_R^c$        &  7/3      & 10/3     \\ \hline 
$L$            & -1      & -2/3      \\ \hline 
$E_R^c$        &  1     & -1/3    \\ \hline
$\lambda$      & -         & -1     \\ \hline
$k    $        & -         & -1     \\ \hline
$\lambda_t$    & -         & -5/3       \\ \hline
$\lambda_b$    & -         & 2/3     \\ \hline
$\lambda_\tau$ & -         & 4        \\ \hline
\end{tabular}
\end{center}
\vspace{0.5cm}
The total superpotential becomes,
\ba  
\label{duperpot}
W &=&\lambda
\frac{\Phi_1^9}{M^9_\P}
N H_{1}H_{2}-
\frac{k}{3}
\frac{\Phi_1^9}{M^9_\P}
N^3 \nonumber\\
& & + \lambda_t\frac{\Phi_1}{M_\P} Q_L H_2 U_R^c+
\lambda_b\frac{\Phi_1^5}{M^5_\P} Q_L H_1 D_R^c+
\lambda_\tau\frac{\Phi_1^4\Phi_2}{M^5_\P} L H_1 E_R^c
\ea  
which is invariant and anomaly free under $U(1)_X$ and the
transformation in eq.(\ref{dualtrans}).  Clearly this example is
not the most elegant of solutions; string models with such
complicated weights and charges have not been derived, and in
addition the higgs couplings are suppressed by a factor of at least
$\rho_\phi^9/M_\P^9$, irrespective of the values one takes for {\em
any} of the free parameters above. For reasonable electroweak
symmetry breaking $\rho_\phi /M_\P$ will not be small.

So at least for the case of $U(1)_X$, this mechanism does not allow
very convincing models to be constructed. The alternative is to embed
the $Z_3$ in the centre of a non-abelian group which could be $SU(3
n)$ or $E_6$. This certainly eases anomaly cancellation although the
presence of the $N^3$ term is particularly problematic since, being
symmetric, it requires large multiplets. Let us finish this section
by considering an example, based on
the $SU(3)_c\times SU(3)_L\times SU(3)_R $ unification (see
ref.\cite{christoph} for a summary of the phenomenology).  

In this model, the particle content is,
\ba
Q &=& (3,3,1) =
\left( 
\begin{array}{c}
          u \\
          d \\
          D 
\end{array} \right)_L \nonumber\\
q &=& (\overline{3},1,\overline{3}) = 
\left( 
\begin{array}{c}
          u \\
          d \\
          D 
\end{array} \right)_R \nonumber\\
L &=& (1,\overline{3},3) = \left( 
\begin{array}{ccc}
          H_2^0 & H_1^- & e_L^- \\
          H_2^+ & H_1^0 & \nu_L \\
          e_R^+ & \nu_R^c & N 
\end{array} \right)_L ,
\ea
where there are three generations, and colour indices on the quarks.
The pattern of symmetry breaking is 
\ba 
SU(3)_c \times SU(3)_L \times SU(3)_R &\rightarrow &
SU(3)_c \times SU(2)_L \times SU(2)_R \times U(1)_{B-L} \nonumber\\
&\rightarrow &
SU(3)_c \times SU(2)_L \times U(1)_Y .
\ea
The hypercharge is given by
\be
Y = T^3_R + \frac{1}{\sqrt{3}}(T^8_L+T^8_R)
\ee
where $T^i $ are the standard Gell-Mann matrices for $SU(3)$. Clearly 
with this assignment of fields, the $Z_3 $ symmetry is simply given by 
a rotation in the 
\[
T^8 = \frac{1}{\sqrt{3}}\left(\begin{array}{ccc}
1 & 0 & 0 \\
0 & 1 & 0 \\
0 & 0 & -2 \end{array} \right)
\]
direction. The superpotential is of the form 
\be
W_{\rm higgs} = \lambda \epsilon^{jln} \epsilon_{ikm} L^i_j L^k_l L^m_n -
 \frac{k}{3}\frac{{\Phi }^{jln}_{ikm} }{M_\P} L^i_j L^k_l L^m_n + 
\lambda_{\rm yuk} L^i_j q^j Q_i .
\ee
The first term is responsible for the $\lambda N H_1 H_2 $ terms as
well as the lepton Yukawa couplings.  The second term is responsible
for the $N^3$ coupling, for which the superfield, $\Phi= 
(1,10,\overline{10})$ had to be introduced.  Since the anomaly
contribution of the $10 $ is 27 times that of the fundamental
representation, the simplest way to cancel anomalies is to include an
additional $\Phi^{'} = (1,\overline{10},10)$ field.

The $\Phi $ field must get VEV along the flat $\Phi_{333}^{333} $
direction, giving the first stage of symmetry breaking. A VEV along
$\Phi_{222}^{'333} $ can give the second.  The third term provides the
Yukawa couplings for the quarks, and also a mass for the $D$-quark
which (through two possible triplet quark couplings which have been
omitted here) would otherwise mediate proton decay.  This requires a
high VEV for at least one of the $N$ fields which, since it
contributes to the first stage of symmetry breaking, must also be in a
flat direction.  With a suitable choice of generation dependence in
the couplings, one can then arrange for the low energy particle
content to be that of the NMSSM.

Finally, the stability of the strings requires that $\pi_1
(G/H) \neq 0 $. Initially the $Z_3 $ symmetry is given by rotations
through $\exp ( 2 \pi i /3 )$ for every $L^i_j$. As long as only
$\Phi^{333}_{333} $ has a VEV, this remains as a global symmetry so
that
\be
\pi_1 (G/H) = \pi_1 \left( \frac{SU(3)_L\times SU(3)_R}{SU(2)_L\times
    SU(2)_R\times U(1)_{B-L}\times Z_3} \right) 
= \pi_0 (H) = Z_3
\ee
since $\pi_1 (G)=\pi_0(G)=0$. (The discrete symmetry may be larger if
there are zeroes or degeneracies in the Yukawa coupling.) 
So one expects $Z_3$ strings to form.
However, a VEV for any of the three generations of $N$ break the $Z_3$
as well. This means that no $N$ should get a VEV until $T\sim M_W $
otherwise the strings would be removed by domain walls before
electroweak symmetry breaking (since the $Z_3$ symmetry that would
remain in the low energy theory would not be embedded in $SU(3)$). For
this one can appeal to the cosmological scenario described earlier,
with all the $N$ VEVs remaining trapped at the origin until $T\sim
m_{3/2}$, when two of them acquire Planck scale VEVs.

These examples show that, at least in principle, it is possible to
make discrete symmetries in models such as the NMSSM innocuous by
embedding them in continuous gauge symmetries at the Planck scale.
However the success or otherwise of this mechanism depends rather
heavily upon the cosmology. Because of this an exact
discrete symmetry at the weak scale still seems to be unlikely, and
models such as the NMSSM remain questionable.

\section{The `Indiscrete' Symmetry Case}

In this section the case where there is no discrete
symmetry at the weak scale is examined. In ref.\cite{us} it was shown that
breaking the discrete symmetry by gravitationally suppressed terms
cannot remove the walls before nucleosynthesis without
destabilising the gauge hierarchy.  This means that 
either $\mu\neq 0$ or both $\mu'\neq 0$ and $k\neq 0$ in addition
to $\lambda\neq 0$. It will be shown  that presently there exist
two ways in which a gauge singlet may be accommodated at the weak
scale without fine-tuning.  These are $R$-symmetry and duality
symmetry in a string effective action. What these symmetries have
in common is that under them one requires different charges for the
superpotential and K\"ahler potential.

First let us recapitulate the arguments of ref.\cite{us}, by considering
adding an ordinary gauge symmetry to the Lagrangian which exists at
the Planck scale. For simplicity, again take this to be a $U(1)_X$
symmetry. The effective couplings at the weak scale are in general
arbitrary functions of hidden sector fields which carry charge under
the new $U(1)_X$ which shall be referred to collectively as $\Phi$ (with
$\xi =\Phi/M_\P$). It is simple to see that one cannot use this
symmetry to forbid terms linear in $N$ thereby avoiding fine-tuning
(again assuming there is an explanation for the smallness of
$\mu$ and $\mu'$ in which all couplings are of order unity).
 
If $\mu (\xi)\neq 0$ then $\mu (\xi)$ must have the same charge as
$\lambda (\xi) N$ and therefore $(\mu (\xi))^\dagger \lambda (\xi) N $
is uncharged. If both $\mu'\neq 0$ and $k\neq 0$ then $\mu' (\xi)$
must have the same charge as $k (\xi) N$ and therefore $(\mu'
(\xi))^\dagger k (\xi) N $ is uncharged. Either way an operator of the
form $f(\xi,\overline{\xi})M_\P N +\hc$ is allowed in the K\"ahler
potential.  Typically $\langle e^{G}\rangle =\langle e^{K} W
\overline{W} \rangle =\ord (m_{3/2}^2) $, so that the terms
\be 
\label{linearN}
m_{3/2} M_\P f(\xi) H_1 H_2 \mbox{ ; } 
m_{3/2} M_\P f(\xi) N^2     \mbox{ ; } 
m_{3/2}^2 M_\P f(\xi) N  
\ee
also appear in the potential, unless one sets $f(\xi)M_\P N =0$
initially.  

It should first be made clear that the 
`explanation for the smallness of $\mu$ and $\mu'$' is here
taken to be   
something like the Giudice-Masiero mechanism~\cite{gm}, in which the
couplings responsible are of order unity. There also exist models in
which the smallness of $\mu$ and $\mu'$ is created by
non-renormalisable couplings of the singlet to fields whose VEVs are
small. In fact in this case things are worse, because although at tree
level the terms in eq.(\ref{linearN}) are indeed small, the singlet
field $N$ couples to the supersymmetry breaking in the hidden
sector through a tadpole diagram. This can be seen from the
logarithmically divergent terms appearing at one loop from~\cite{destab}
\be
V_{1-loop} = \frac{\log \Lambda^2}{32 \pi^2}
 \int \mbox{d}^4\theta e^{2 K/3 M_\P^2} \varphi\overline{\varphi}
W_{ij} \overline{W}^{ij} +\ldots
\ee
where here the indices denote differention which is covariant with
respect to the K\"ahler manifold and $\varphi $ is the chiral
compensator. (Supersymmetry breaking is embodied in the $\theta $
dependence in the VEVs of $K$ and $\varphi$.)
These are the divergent terms which lead to logarithmic
running of the soft-breaking scalar masses. However, if there is a 
$\mu$-term produced directly in the superpotential from some product
of hidden sector fields ($\mu = \Phi^m/M_\P^{m-1}$ for
example), the contribution above includes
\be
 \frac{\log \Lambda^2}{32 \pi^2}
 \int \mbox{d}^4 \theta \mu (\Phi) \lambda^\dagger N^\dagger = 
 \frac{\log \Lambda^2}{32 \pi^2} \lambda^\dagger F_N^\dagger  
\frac{m \phi^{m-1} F_\Phi}{M^{m-1}_\P}
\sim \left( \frac{M_\P}{M_W}\right)^{1/m}M_W^{2} F_N^\dagger . 
\ee
where since $\Phi$ is a hidden sector field, it is assumed that 
$F_\Phi \sim M_W M_\P $, and that also  
$\langle |\phi |^m \rangle \sim M_W M_\P^{m-1}$ in
order to get $\mu \sim M_W $. This leads to a value of $F_N \gg M_W$
unless $m$ is extremely large, destabilising the gauge hierarchy.

One may easily demonstrate that the above arguments also apply when
the additional group is non-abelian.  For example consider extending
the $SU(3)^3 $ example of the previous section by adding extra $\mu$
and $\mu'$ terms in the K\"ahler potential. The $\mu' N^2 $ term may
be generated from the coupling $\Theta_{ik}^{\dagger jl} L^i_j L^k_l
/M_\P $ where $\Theta = (1,\overline{6},6)$ and $\Theta^{33}_{33}$
acquires a VEV.  But now the linear term $(\Theta \Phi )^j_i L^i_j =
(\Theta \Phi )^3_3 N $ is also allowed.  Similarly, the $\mu H_1 H_2 $
term requires $(\epsilon\epsilon\Theta^\dagger )_{ik}^{jl} L^i_j L^k_l
/M_\P $ where $\Theta = (1,3,\overline{3})$ and $\Theta^{3}_{3}$
acquires a VEV.  This then allows the coupling $\Theta_i^j L_j^i=
\Theta^3_3 N $.  (In terms of the component fields, in this case the
low energy singlet $N$ mixes with singlets which get a VEV of
$\cal{O}(M_\P)$.)

One should bear in mind that these terms do not necessarily lead to a
destabilisation of the gauge hierarchy since in most (but not all)
cases, if one sets the linear coupling to zero in the first place, it
remains small to higher order in perturbation theory.  So this is 
merely a fine-tuning problem. One might also argue that this 
fine-tuning problem is of a less serious nature than the
$\mu$-problem, since in the latter the coupling has to be very
small, whereas here the coupling may just happen to be absent 
(as for example are superpotential mass terms in string theory). 
However, in addition to the simplest 
operators, there are generally many more (possibly an infinite 
number of) operators which must be set to zero.

This becomes evident when one continues with the $U(1)_X$ case. 
Take for example the K\"ahler function in ref.\cite{us};
\be 
{\cal G} = z^i z^\dagger_i + \Phi \Phi^\dagger 
+ \Phi ' \Phi ^{'\dagger}  
+ \left( \frac{\alpha}{M_\P}\Phi^{' \dagger}H_1 H_2 + 
 \frac{\alpha '}{M_\P}\Phi^{' \dagger}N^2 +\hc \right) 
+ \log |h(z,\Phi ) + g(\Phi , \Phi' )|^2 ,
\ee
where again, $z^i$ are the visible sector fields, and $h(z ,\Phi )$ is the 
superpotential of the NMSSM given in eq.(\ref{nmssmpot}) with charge
assignments as before. Under the $U(1)_X$, $\Phi ' $ must have charge 2 and 
so the operator 
\be
\Phi' \Phi N 
\ee
is also charge invariant and must be set to zero. Clearly there is a
very large number of additional, nonrenormalisable operators $f(\Phi ,
\Phi ') N$ which should not appear (assuming that $\langle \Phi
\rangle = {\cal O}(M_\P)$) as well as $f(\Phi ,\Phi ') \Phi^2 N^2 $
and $f(\Phi ,\Phi ') \Phi^2 H_1 H_2 $. (Here $f$ implies general
polynomials with the correct charge, for example $\Phi^2 \Phi'^3$.)
These are the operators which directly destabilise the gauge
hierarchy. {\em In addition} there are operators which destabilise the
hierachy through divergences; for example at one loop order the
potential receives contributions of the form $ \Lambda^2 R^k_n (W
e^K)_k (\overline{W} e^K)^n $, where $R^k_n$ is the Ricci tensor for
the K\"ahler manifold, and $k$ and $n$ subscripts denote
differentiation with respect to the fields (raising and lowering of
indices being done by $K_n^{\overline{l}} $)~\cite{destab}. Jain in 
ref.\cite{destab} has
shown that destabilising divergences occur for any couplings of the
form $f(\Phi ,\Phi ') z z z^\dagger $. This means that the operators
$f(\Phi , \Phi ') N H_i H^{\dagger}_i$ and $f(\Phi , \Phi ') N N
N^{\dagger}$ are also disallowed. Progressing to higher loop order,
the operators $\Phi^4 \Phi '^4 N^4 $, $\Phi^4 \Phi '^4 N^2 H_1 H_2 $
and $\Phi^4 \Phi '^4 (H_1 H_2)^2 $ appearing in the {\it
superpotential} destabilise the gauge hierarchy through two and three
loop diagrams, and so on. Obviously the degree of fine-tuning
decreases with higher order since each loop gives a factor $\Lambda^2
/(16 \pi^2 )$  where $\Lambda $ is a cut-off, and involves more Yukawa
couplings. It therefore seems reasonable to assume that contributions 
which are higher than six-loop 
are unable to destabilise the hierarchy. Upto and including six loop, 
the following operators could be dangerous if they
appear in the superpotential (multiplied by any function of 
$f(\Phi, \Phi'))$, since one can write down a tadpole diagram using
them (together with the trilinear operators of the NMSSM);
\vspace{0.5cm}
\begin{center}
\begin{tabular}{||l|r|r||}   \hline
  \mbox{Operator}     & \mbox{Loop-order of diagram} \\ \hline\hline 
    $N^2$, $H_1 H_2$     & 1         \\ \hline 
    $N^4$, $N^2 H_1 H_2$     & 2         \\ \hline 
         $(H_1 H_2)^2  $, $N (H_1 H_2)^2$,
$N^3 (H_1 H_2)$, $N^5$     & 3         \\ \hline

$N^3 (H_1 H_2)^2$, $N^5 (H_1 H_2)$, $N^7$    & 4         \\ \hline
$N (H_1 H_2)^3$, $N^2 (H_1 H_2)^3$, $N^4 (H_1 H_2)^2$,
$N^6 (H_1 H_2)$, $N^8$     & 5         \\ \hline
$N^2 (H_1 H_2)^4$, $N^4 (H_1 H_2)^3$, $N^6 (H_1 H_2)^2$,
$N^8 (H_1 H_2)$, $N^{10}$     & 6         \\ \hline
\end{tabular}
\end{center}
\vspace{0.5cm}
Notice that, since the leading divergences involve chiral or antichiral
vertices only, an operator must break the $Z_3$ symmetry in 
$h(z,\Phi)$ in order for it to be dangerous (so that for example 
$N^2 (H_1 H_2)^2 $ does not destabilise the hierarchy). 
These are all the operators which lead to a superspace tadpole
diagram, but it turns out that only those which have {\em even}
dimension are dangerous. This can be seen as follows. 

First, using the supergraph rules described in refs.\cite{destab,sriv}
for the leading rigid supergraphs, it may easily be seen that only
diagrams with an even number of vertices are dangerous. A chiral
vertex, $A$, with $L_A$ internal legs, throws $L_A-1$ of the
${\overline D}^2$ operators onto the surrounding propagators. These
may be manipulated in the standard manner by partial differentiation
to expose the $\delta (\theta - \theta')$ function on a propagator,
allowing the integrations over $\theta $s to be carried out ($\theta $
and $\theta'$ belong to the vertex at either end of the
propagator). Products of three or more $D^2 $ may be reduced using the
identities $D^2 \overline{D}^2 D^2 = 16 \partial^2 D^2 $ and
$\overline{D}^2 D^2 \overline{D}^2 = 16 \partial^2 \overline{D}^2
$. Pairs of $D^2$ operators are also removed on integration over
$\theta $s, since
\be
\int {\rm d}^4 \theta' \delta  (\theta -
  \theta') D^2 \overline{D}^2 \delta  (\theta -
  \theta') =16.  
\ee
A single $D^2$ may also be removed by acting on the
$(e^{K/3}/\phi\overline{\phi})_{\rm classical}$ factors on each
propagator~\cite{destab}, but this renders the diagram innocuous. 
The end result is an integral over a single $ \theta $,
but clearly only if the initial total number of $D^2 $ and $\overline{D}^2$ 
operators was even. This number is given by 
\be
\sum_A^V (L_A-1) = 2 P - V = {\rm even} 
\ee   
where $P$ is the number of propagators, and $V$ is the number of
vertices; hence the number of vertices should be even. 

Now consider constructing a tadpole diagram beginning with a single
vertex with an odd number of legs. When more vertices are added, if the
extra non-renormalisable operator has an odd number of superfields
then there are only vertices with odd numbers of legs to choose from. 
Adding a single vertex with an odd number of legs changes
the total number of external legs by an odd number. So in order to
have a tadpole diagram one has to add an even number of vertices
implying that $V$ is odd, and that the diagram is therefore
harmless. Hence only operators with even numbers of superfields can be
harmful to the gauge hierarchy. Counting in addition the $N$ operator 
itself, this means that in this case 17 operators (multiplied by any 
appropriately charged function $f(\Phi, \Phi')$) must be set to 
zero by hand.
\\

The reason that it has not been possible to forbid divergences linear in $N$ 
in the models that have been discussed here and in ref.\cite{us}, is
that the K\"ahler potential and superpotential have the same charges
(i.e. zero).  There are however two available symmetries in which the
K\"ahler and superpotentials transform differently. These may
accommodate singlet extensions to the MSSM simply and without fine-tuning.

The first is gauged $U(1)_R$-symmetry~\cite{herbi,gaugedr}. In this case the 
K\"ahler potential has zero $R$-charge, but the superpotential has $R$-charge 
2. This means that the standard renormalisable NMSSM higgs superpotential,
\be 
\label{rwhiggs}
W_{\rm higgs}=\lambda N H_{1}H_{2}-
\frac{k}{3}N^3,
\ee
has the correct $R$-charge if $R(N)= 2/3$ and $R(H_1)+R(H_2) =4/3 $. 
So consider the K\"ahler potential   
\be 
\label{quad}
{\cal G} = z^i z^\dagger_i + \Phi \Phi^\dagger 
+ \Phi ' \Phi ^{'\dagger}  
+ \left( \frac{\alpha}{M^2_\P}\Phi\Phi^{' \dagger}H_1 H_2 + 
 \frac{\alpha '}{M^2_\P}\Phi\Phi^{' \dagger}N^2 +\hc \right) 
+ \log |h(z ) + g(\Phi , \Phi' )|^2 ,
\ee
where $h(z)$ is the superpotential involving just visible sector fields 
and $\Phi$, $\Phi'$ again represent hidden sector fields with superpotential 
$g(\Phi , \Phi')$ (they may represent arbitrary functions of hidden 
sector fields in what follows). Both $\Phi $ and $\Phi^{'\dagger}$ appear 
here in order to prevent unwanted couplings being allowed in the 
superpotential which must be a holomorphic function of superfields.

Invariance of the K\"ahler potential requires that $R(\Phi
)+R(\Phi^{'\dagger}) =4/3 $. If all the $R$-charges are chosen to be
positive, then the terms in eq.(\ref{rwhiggs}) are clearly the only
functions which can appear in the superpotential (since all higher
dimension ones have $R$-charge greater than 2).  (The $R$-charges of
$\Phi $ and $\Phi'$ must of course be chosen to be sufficiently
obtuse; for example $R(\Phi )= 16/3 $, $R(\Phi ')= 4$, is adequate,
because the lowest $R$-charge one can make with them is $\pm 4/3$.)
Moreover it is easy to see that with this set of $R$-charges there can
never be a coupling which is linear or trilinear appearing in the
K\"ahler potential.  In fact the operators
\be
\hat{O}_{mkl}=
f N^{2m + 2k - 2l} (H_1 H_2 )^l (H_2^\dagger
H_1^\dagger)^k 
\ee
where $f$ is an arbitrary $R$-invariant function, satisfy 
\be
- m \left( R(\Phi ) + R (\Phi^{'\dagger})\right) =
R ( N^{2 m} ) = R(\hat{O}_{mkl} ).
\ee
The quadratic coupling in eq.(\ref{quad}) results when $m=1$, and when
$m=2$, the couplings $N^4$, $N^2 (H_1 H_2) $ and $(H_1 H_2)^2$,
but this time appearing only in the K\"ahler potential, not the
superpotential. Such couplings do not destabilise the hierarchy.  In
particular the dangerous trilinear terms in the K\"ahler potential
have been avoided. 

The second symmetry one can use to forbid terms linear in $N$ is
target space duality. Consider for example the K\"ahler potential
defined in eq.(\ref{stringkahler}), but now, as in
refs.\cite{moduli2,brignole}, identifying $\Phi_1$ and $\Phi_2$ with
the higgs superfields $H_1$ and $H_2$ both of which have weight
$-1$. This identification generates a $\mu H_1 H_2 $ term in the
low-energy lagrangian~\cite{moduli2,ant1,brignole}.  The remaining
dependence on the matter fields may be written as
\be
\label{kahler2}
\delta K_{\rm matter}= K_i^k z^i z_j^\dagger +\ldots 
\ee
and is invariant under the modular transformation in eq.(\ref{dualtrans}).  
Here the ellipsis represents terms higher order in the expansion. 
Since the superpotential must transform as in eq.(\ref{wdual}), one
can consider the usual NMSSM superpotential if 
\ba
3 n_N + n_k &=& -1 \nonumber\\
n_N + n_\lambda &=& +1,
\ea
which, taking $n_k=0$, gives $n_N=-1/3$ and $n_\lambda = 4/3$.  
 
With this assignment of duality charges the first extra term which 
can appear in the superpotential is the seventh order $\lambda^3 N^3
(H_1 H_2)^2 /M_P^4 $ operator. Since this operator has an odd number 
of fields it cannot destabilise the hierarchy by itself for reasons
discussed earlier. Furthermore {\em only} odd operators can 
result with this choice of weights, since the required weight
$-1 $ is an odd multiple of $n_N$ whilst those of $(H_1 H_2)$ and 
$\lambda $ are both even multiples of $n_N$. 

As for the K\"ahler potential, one 
expects the extra terms in the expansion of eq.(\ref{kahler2}) to be 
multiplied by powers of $(T+\overline{T})$. Thus terms in which the
holomorphic and anti-holomorphic weights are the same may be allowed.
The $\lambda N^2 (H_1 H_2) $ coupling (which in fact has weight zero)
is the lowest dimension operator which satisfies this criterion.

There are clearly many ways in which one could devise similar models.
A perhaps more obvious example would be models in which the
superpotential transforms with weight $-3$. There all the physical
fields could be given weight $-1$, with the couplings having weight
$0$. It is then clear that only trilinear couplings can exist in
the superpotential, and only even-dimension terms can appear in the
K\"ahler potential.
  
\section{Conclusions} 

The possibilities for extending the MSSM by a
singlet field have been examined, and constraints from fine-tuning,
primordial nucleosynthesis and cosmological domain walls have been 
applied. The most appealing models have no discrete or continuous 
global symmetries at the weak scale.

For the case of continuous global symmetry, the existence of an
axion, and the cosmological bounds associated with it, make a
reasonable phenomenology difficult to achieve without
fine-tuning. At the very least, one can rule out the Peccei-Quinn
mechanism here, unless the continuous symmetry is broken to a high
order, discrete symmetry by gravitationally suppressed terms. For
this to be successful additional singlets are required.

When there is a discrete symmetry at the electroweak scale,
breaking it by gravitationally suppressed terms cannot remove the
associated domain walls before nucleosynthesis, without reintroducing
the hierarchy problem. A suitable solution
to the ensuing domain wall problem may be to embed the discrete
symmetry in a gauge symmetry.  Although the cosmology is then
necessarily rather complicated, it seems that, at least for the
$SU(3)^3$ model we discussed, the emerging picture of string
cosmology may be able to accommodate it.

The most efficient way to accommodate the singlet is however to
impose a gauged $R$-symmetry or a target space duality on the full
theory including gravity. In this case one expects all couplings
(i.e. $\mu H_1 H_2 $, $\mu N^2 $, $\lambda N H_1 H_2 $ and $k N^3
$) to be possible in the weak scale effective theory. The phenomenological 
implications of these more general cases, have been discussed recently 
in ref.\cite{moorehouse}. 

\vspace{1cm}
\noindent
{\bf \Large Acknowledgements:} I would like to thank H.~Dreiner,
J.-M.~Fr\`ere, M.~Hindmarsh, S.~King, D.~Lyth, G.~Ross, S.~Sarkar 
and P.~Van Driel for valuable discussions. This work was 
supported in part by the European Network Flavourdynamics
(ref.chrx-ct93-0132) and by INTAS project 94/2352.


\newpage
\small
\begin{thebibliography}{99}
%
\bibitem{mssm}
 For reviews see, H.~P.~Nilles, Phys.~Rep.~110 (1984) 1;
 H.~E.~Haber and G.~L.~Kane, Phys.~Rep.~117 (1985) 75.
%
\bibitem{muprob}
 L.~Hall, J.~Lykken and S.~Weinberg, Phys.~Rev.~D27 (1983) 2359;
 J.~E.~Kim and H.~P.~Nilles, Phys.~Lett.~B138 (1984) 150;
 K.~Inoue, A.~Kakuto and T.~Takano, Prog.~Theor.~Phys.~75 (1986) 664;
 A.~A.~Ansel'm and A.~A.~Johansen, Phys.~Lett.~B200 (1988) 331.
%
\bibitem{gm}
 G.~Giudice and A.~Masiero, Phys.~Lett.~B206 (1988) 480
%
\bibitem{aw}
 S.~A.~Abel and P.~L.~White, Phys.~Rev.~D52 (1995) 4371.
%
\bibitem{us}
 S.~A.~Abel, S.~Sarkar and P.~L.~White, Nucl.~Phys.~B454 (1995) 663.
%
\bibitem{graham}
C.~A.~Casas and G.~G.~Ross, Phys.~Lett.~B192 (1987) 119; 
\ibid B198 (1987) 461.
%
\bibitem{nmssm}
 P.~Fayet, Nucl.~Phys.~B90 (1975) 104;
 H.-P.~Nilles, M.~Srednicki and D.~Wyler, Phys.~Lett.~B120 (1983) 346;
 J.-M.~Frere, D.~R.~T.~Jones and S.~Raby, Nucl.~Phys.~B222 (1983) 11;
 J.-P.~Derendinger and C.~A.~Savoy, Nucl.~Phys.~B237 (1984) 307;
 B.~R.~Greene and P.~J.~Miron, Phys.~Lett.~B168 (1986) 226;
 L.~Durand and J.~L.~Lopez, Phys.~Lett.~B217 (1989) 463;
 M.~Drees, Intern.~J.~Mod.~Phys.~A4 (1989) 3645;
 J.~Ellis, J.~Gunion, H.~Haber, L.~Roszkowski and F.~Zwirner,
  Phys.~Rev.~D39 (1989) 844;
 U.~Ellwanger, M.~Rausch~de~Traubenberg and C.~A.~Savoy,
  Phys.~Lett.~B315 (1993) 331; Z.~Phys.~C67 (1995) 665;
 T.~Elliott, S.~F.~King and P.~L.~White, Phys.~Lett.~B351 (1995) 213;
 S.~F.~King and P.~L.~White, Phys.~Rev.~D52 (1995) 4183.
%
\bibitem{ellis}
 J.~Ellis, K.~Enqvist, D.~V.~Nanopoulos, K.~Olive, M.~Quiros and
  F.~Zwirner, Phys.~Lett.~B176 (1986) 403.
%
\bibitem{ls}
 G.~Lazarides and Q.~Shafi, Phys.~Lett.~B115 (1982) 21;
 T.~W.~B.~Kibble, G.~Lazarides and Q.~Shafi, Phys.~Rev.~D26 (1982)
 435; Phys.~Lett.~B113 (1982) 237;
 S.~M.~Barr, D.~B.~Reiss and A.~Zee, Phys.~Lett.~B116 (1982) 227.
%
\bibitem{destab}
 U.~Ellwanger, Phys.~Lett.~B133 (1983) 187;
 J.~Bagger and E.~Poppitz, Phys.~Rev.~Lett. 71 (1993) 2380;
 J.~Bagger, E.~Poppitz and L.~Randall, Nucl.~Phys.~B455 (1995) 59;
 V.~Jain,  Phys.~Lett.~B351 (1995) 481. 
%
\bibitem{zko}
 Ya.~B.~Zel'dovich, I.~Yu.~Kobzarev and L.~B.~Okun, 
 Sov.~Phys.~JETP 40 (1975) 1.
%
\bibitem{rai}
 B.~Rai and G.~Senjanovic, Phys.~Rev.~D49 (1994) 2729.
%
\bibitem{herbi}
 A.~H.~Chamseddine and H.~K.~Dreiner, Nucl.~Phys.~B458 (1996) 65.
%
\bibitem{gaugedr}
 D.~Z.~Freedman, Phys.~Rev.~D15 (1977) 1173;
 A.~Das, M.~Fischler and M.~Rocek, Phys.~Rev.~D16 (1977) 3427;
 B.~de~Wit and P.~van~Nieuwenhuizen, Nucl.~Phys.~B139 (1978) 216; 
 K.~S.~Stelle and P.~C.~West, Nucl.~Phys.~B145 (1978) 175;
 R.~Barbieri, S.~Ferrara, D.~V.~Nanopoulos and K.~S.~Stelle,
 Phys.~Lett.~B113 (1982) 219;
 S.~Ferrara, L.~Girardello, T.~Kugo and A.~van~Proeyen,
 Nucl.~Phys.~B223 (1983) 191;
 D.~J.~Casta\~no, D.~Z.~Freedman and C.~Manuel, MIT-CTP-2454, 
 hep-ph/9507397.
%
\bibitem{DFSZ}
A.~R.~Zhitnisky, Sov.~J.~Nucl.~Phys.~31 (1980) 260;
M.~Dine, W.~Fischler and M.~Srednicki, Phys.~Lett.~B104 (1981) 199.
%
\bibitem{CPrefs}
See for example, J.~E.~Kim, Phys.~Rep.~150 (1987) 1;
 H.-Y.~Cheng,  Phys.~Rep.~158 (1988) 1; 
 G.~Raffelt, Phys.~Rep.~198 (1990) 1.
%
%
\bibitem{wallevol}
 A.~Vilenkin and A.~E.~Everett, Phys.~Rev.~Lett.~48 (1982) 1867;
 A.~E.~Everett and A.~Vilenkin, Nucl.~Phys.~B207 (1982) 43;
 W.~H.~Press, B.~S.~Ryden and D.~N.~Spergel,
  Astrophys.~J.~347 (1989) 590; \ibid 357 (1990) 293;
 A.~Vilenkin and E.~P.~S.~Shellard, {\em `Cosmic Strings and Other
  Topological Defects'} (Cambridge University Press, 1994).
%
\bibitem{cosmobounds}
 J.~Ellis \etal, Nucl.~Phys.~B373 (1992) 399;
 S.~Sarkar, Oxford preprint OUTP-95-16P, hep-ph/9602260. 
%
\bibitem{duality}
 For a review see, A.~Giveon, M.~Porrati and E.~Rabinovici,
 Phys.~Rep.~244 (1994) 77.
%
\bibitem{david1} 
 L.~Randall and S.~Thomas, Nucl.~Phys.~B449 (1995) 229;
 T.~Banks \etal, Phys.~Rev.~D52 (1995) 75; T.~Banks, 
 M.~Berkooz and P.~J.~Steinhardt, Phys.~Rev.~D52 (1995) 3548;
 D.~H.~Lyth and E.~D.~Stewart, Phys.~Rev.~Lett.~75 (1995) 201; 
 T.~Barreiro, E.~J.~Copeland, D.~H.~Lyth, T.~Prokopec, LANCASTER-TH-95-07, 
hep-ph/9602263. 
%
\bibitem{david2}
 D.~H.~Lyth and E.~D.~Stewart, Phys.~Rev.~D51 (1996) 1784
%
\bibitem{subir}
 G.~D.~Coughlan \etal, Phys.~Lett.~B131 (1983) 59;
 G.~G.~Ross and S.~Sarkar, Nucl.~Phys.~B461 (1996) 597. 
%
\bibitem{moduli}
 E.~Witten, Phys.~Lett.~B155 (1985) 151;
 S.~Ferrara, C.~Kounnas and M.~Porrati, Phys.~Lett.~B181 (1986) 263;
 S.~Ferrara, L.~Girardello, C.~Kounnas and M.~Porrati,
 Phys.~Lett.~B192 (1987) 386; \ibid B194 (1987) 358;
 M.~Cvetic, J.~Louis and B.~Ovrut, Phys.~Lett.~B206 (1988) 227;
 M.~Cvetic, J.~Molera and B.~Ovrut, Phys.~Rev.~D40 (1989) 1140;
 L.~Ib\'a\~nez, H.~P.~Nilles and F.~Quevedo, Phys.~Lett.~B192 (1987)
 332;
 L.~Ib\'a\~nez J.~Mas, H.~P.~Nilles and F.~Quevedo, Nucl.~Phys.~B301
 (1988) 157;
 L.~Ib\'a\~nez and D.~L\"ust, Nucl.~Phys.~B382 (1992) 305;
 J.~L.~Lopez, D.~V.~Nanopoulos, K.~Yuan, Phys.~Rev.~D50 (1994) 4060. 
%
\bibitem{moduli2}
 G.~L.~Cardosos, D.~L\"ust and T.~Mohaupt, Nucl.~Phys.~B432 (1994) 68;
 HUB-IEP-94-18, hep-th/9409095.
%
\bibitem{ant1}
 I.~Antoniadis, E.~Gava, K.~S.~Narain and T.~R.~Taylor, 
 Nucl.~Phys.~B432 (1994) 187.
%
\bibitem{brignole}
 A.~Brignole and F.~Zwirner, Phys.~Lett.~B342 (1995) 117. 
%
\bibitem{kpz}
 C.~Kounnas,  I.~Pavel and F.~Zwirner, Phys.~Lett.~B335 (1994) 403.
%
\bibitem{binetruy}
 P.~Binetruy and E.~Dudas, Nucl.~Phys.~B442 (1995) 21; 
 Nucl.~Phys.~B451 (1995) 31;
%
\bibitem{noscale}
 E.~Cremmer, S.~Ferrara, C.~Kounnas and D.~V.~Nanopoulos, 
 Phys.~Lett.~B133 (1983) 61;
 J.~Ellis, A.~B.~Lahanas, D.~V.Nanopoulos and K.~Tamvakis,
 Phys.~Lett.~B134 (1984) 429;
 J.~Ellis, C.~Kounnas and D.~V.~Nanopoulos, Nucl.~Phys.~B241 (1984)
 406; B247 (1984) 373;
 S.~Ferrara, C.~Kounnas and F.~Zwirner, Nucl.~Phys.~B429 (1994) 589;
 For review see, A.~B.~Lahanas and D.~V.~Nanopoulos, Phys.~Rep.~145
 (1987) 1.
%
\bibitem{dps}
E.~Dudas, S.~Pokorski and C.~A.~Savoy, Phys.~Lett.~B356 (1995) 45.
%
\bibitem{greenschwarz}
M.~Green and J.~Schwarz, Phys.~Lett.~B149 (1984) 117.
%
\bibitem{dsw}
E.~Witten, Phys.~Lett.~B149 (1984) 351;
W.~Lerche, B.~Nilsson and A.~N.~Schellekens Nucl.~Phys.~B299 (1988) 91;
M.~Dine, N.~Seiberg and E.~Witten, Nucl.~Phys.~B289 (1987) 585;
J.~Atick, L.~Dixon and A.~Sen, Nucl.~Phys.~B292 (1987) 109.
%
\bibitem{christoph} 
G.~Lazarides, C.~Panagiotakopoulos and Q.~Shafi, Phys.~Lett.~B315
(1993) 325;
G.~Dvali and Q.~Shafi, Phys.~Lett.~B326 (1994) 258;
C.~M.~A.~Scheich and M.~G.~Schmidt, Int.~J.~Mod.~Phys.~A7 (1992) 8021.
%
\bibitem{sriv}
M.~T.~Grisaru, W.~Siegel and M.~Rocek, Nucl.~Phys.~B159 (1979) 429; 
For reviews see S.~J.~Gates, 
M.~T.~Grisaru, M.~Rocek and W.~Siegel, {\em `Superspace or 1001
  Lessons in Supersymmetry'}, (Benjamin/Cummings, 1983); 
P.~P.~Srivastava, {\em `Supersymmetry, Superfields and
  Supergravity'}, (Adam Hilge, 1986).
%
\bibitem{moorehouse}
A.~T.~Davies, C.~D.~Froggatt and R.~G.~Moorhouse, hep-ph/9603388
%
\end{thebibliography}
\end{document}